\documentclass[aps,prb,twocolumn]{revtex4}
\usepackage{lmodern}
\usepackage[T1]{fontenc}
\usepackage[utf8]{inputenc}
\usepackage{mathtools}

\usepackage{color, amsmath, amsthm, amsfonts, bbm}
\usepackage[pdftex]{graphicx}
\usepackage{epstopdf}
\usepackage{subfig}

\newcommand{\HH}{\mathcal{H}}
\newcommand{\A}{\mathcal{A}}
\newcommand{\Tr}{\operatorname{Tr}}
\renewcommand{\phi}{\varphi}

\newcommand{\ket}[1]{|{#1}\rangle}

\DeclareMathOperator{\sech}{sech}
\newcommand{\bi}{\bibitem}

\newcommand{\ct}{\cite}

\begin{document}
\title{Topological footprints of the 1D Kitaev chain with long range superconducting pairings at a finite temperature}
\author{Utso Bhattacharya and Amit Dutta\\Department of Physics, Indian Institute of Technology, Kanpur-208016, India}

\vspace{-20pt}
\begin{abstract}
We study the 1D Kitaev chain with long range superconductive pairing terms at a finite temperature  \ {where the system is prepared  in a mixed state in
equilibrium with a  heat reservoir maintained  at a constant temperature $T$}. In order to probe the footprint of the ground state  topological behavior of the model at finite temperature, we look at two global quantities extracted out of two  geometrical constructions: the Uhlmann and the interferometric phase. Interestingly, \  {when
the long-range effect dominates}, the Uhlmann phase approach fails to reproduce the topological aspects of the model in the pure state limit;  on the other hand, the interferometric phase, though has a proper pure state reduction, shows a behaviour independent
of the ambient temperature.
\end{abstract}
\maketitle

\section{ Introduction}

A tremendous effort is now being focused at the experimental front to realize topological superconductors which constitute an essential component of quantum computers and simulators. What makes the realization of a topological superconductor an absolute experimental necessity is that it hosts exotic Majorana modes (MMs) as zero-energy localized modes at its edges or boundaries. Such modes are absent in conventional nontopological superconductors. These exotic MMs are topologically protected against local perturbations and cannot be removed unless a global change in the ground state properties in the form of a topological phase transition occurs. This robustness serves as a key property which enables them to be used as qubits to store and manipulate quantum information in a topological quantum computer without the chance of quick loss of information through decoherence. MMs have been proposed to exist in many systems like heterostructures of topological
insulators and s-wave superconductors \cite{fu08}, cold fermion
systems with Rashba spin orbit coupling, Zeeman field, and
an attractive s-wave interaction \cite{zhang08,sato09}, and also, heterostructures
of spin-orbit coupled semiconductor thin films \cite{sau10a,sau10b} or
nanowires \cite{sau10b,lutchyn10,oreg10} proximity coupled with s-wave superconductors
and a Zeeman field. Although, there have been also been claims of observation of MMs in a few experiments  \cite{mourik12,rokhinson12,deng12,das12,churchill13,finck13,alicea12,leijnse12,beenakker13,stanescu13}, they have as yet remained experimentally elusive.\\


On the other hand,   recent  experimental realisation of long-range interacting quantum models with tunable long-range interactions (or long-range pairing term) \cite{expt_longrange} has  renewed the interest in studying the  equilibrium behaviour as well as the non-equilibrium dynamics of quantum models with infinite-range interactions with interaction strength between two sites separated by a distance $r$  falling off in a power-law fashion as $1/r^\alpha$  \cite{vodola14,vodola16,viyuela16,lepori17,regemortel,silva_longrange,fey2016,spin_model_dpt,kzml,dellanna17,unpub,halimeh17,halimeh171,homrighausen17,anirban_dutta17}.
 Let us recall that  a power-law interacting ferromagnetic Ising chain is being studied for more than last four decades \ct{ruelles68,dyson69,dysonn69,kac69,thouless69,yuval71,fisher72,kosterlitz76,cardy81,bhattacharjee81,bhattacharjee82,imbrie88,luijten01};   quantum phase transitions \ct{sachdev11,suzuki13,dutta15} in the corresponding
  quantum  Ising chain with interaction decaying in a power-law fashion was also explored long-back \ct{dutta01}.

 Recently motivated by  the short-range  one dimensional Kitaev chain \ct{kitaev01},  a long-range version of  an {\it integrable} $p$-wave superconducting chain of fermions, with a long-range
super-conducting pairing/gap term  has been proposed \cite{vodola14,vodola16,viyuela16};  interestingly, in this model the $2 \times 2$ structure corresponding to each momentum value survives in spite of the power-law interacting super-conducting term. It has been observed \cite{viyuela16}  that when the pairng terms decay faster, the model captures short-range topological superconducting physics; on the contrary,  for slow decay of the long range interactions given by $\alpha<1$, the model supports a new unconventional topological phase of matter. In this new phase,  the zero energy MMs coalesce to form massive nonlocal edge states called massive Dirac Modes which are otherwise absent in the standard Kitaev model. These new edge states lie within the bulk energy gap and are topologically protected against local perturbations that do not break fermionic parity and particle-hole symmetry and may eventually find novel applications in the field of topological quantum computations.

Although, throughout the last century, phases of matter have been very successfully characterised by 
taking recourse to a local order parameter in accordance with Landau's theory, the 
order parameter required to classify such 1D topological superconductors studied here are however global in nature. Indeed at zero temperature, most of the phase diagram for the 1D Kitaev chain with long range pairings  can be understood using the conventional winding number used to classify a standard 1D p-wave topological superconductor. However, an intriguing question remains as to what extent the topological properties of such a long range paired system would survive when coupled to a heat reservoir at some constant temperature $T$. We here note that Viyuela $et~ al.$,  \ct{viyuela14} (see also \ct{huang14}) introduced the Uhlmann geometric phase \cite{uhlmann86,uhlmann89} as a tool to characterize symmetry-protected topological phases in 1D fermion systems described by a  Gibbs' ensemble. They illustrated that not only the Uhlmann phase acts as a global order parameter which can classify the two different topological phases in the standard 1D Kitaev chain but thay also demonstrated that there exists a critical temperature at which the Uhlmann phase goes discontinuously and abruptly to zero. Furthermore, at small temperatures, they showed that the Uhlmann phase can also capture the expected behavior of topological phase in such fermionic systems. Subsequently, the behaviour of a different geometric phase, introduced in the context of interferometry by Sjoqvist $et~ al.$ \ct{sjoqvist00}, has also been studied in the context of the short-range 1D Kitaev chain which shows contrasting behavior to that of the Uhlmann approach \ct{andersson16}. (For a review on these two approaches, see [\onlinecite{zela12}]. We note in passing that recently the interferometric phase approach have also been found to be 
relevant in the context of mixed state dynamical quantum phase transitions \ct{bhattacharya17,heyl17} and also in the context of mixed state topology \ct{bardyn18}. \\

In this work, we therefore, consider a 1D Kitaev chain with a long range superconducting pairing term after it has thermalised by being in contact with a heat reservoir at temperature $T$ and is effectively described by a  Gibbs' ensemble. In order to probe the topological aspects of the model considered, two disjoint approaches are pursued, namely the Uhlmann geometric approach and the interferometric geometric approach. The two main quetions addressed here through the two approaches are the following: (a) Can both the approach properly reproduce the topological phase diagram in the pure state (or the zero temperature) limit? (b) What do the two approaches reveal about the  extent of the survival of the topological properties in this long range superconducting scenario?  

The paper is organised in the following fashion: in Sec. \ref{LRK}, we review the topological phase diagram of long-range Kitaev (LRK) chain. In Secs. \ref{UA} and \ref{IGP},
the LRK chain is studied at a finite temperature using Uhlmann phase and interferometric phase approaches, respectively. Concluding comments are presented
in Sec. \ref{sec_conclusion}.

\section{The Long Range Interacting Kitaev chain}\label{LRK}

Let us consider a simple model of spinless fermions on a 1-D lattice with long range p-wave superconducting pairings, known as the long-range interacting Kitaev (LRK) chain. The Hamiltonian is of the form \cite{vodola14}:
 
\begin{equation}
\begin{split}
H & = \sum_{n=1}^{N} \Big\{ - t\left(c_{n+1}^\dagger c_{n} + c_{n}^\dagger c_{n+1}\right) - \mu 
c_n^\dagger c_n  \\
& +\sum_{l=1}^{N-1} \frac{\Delta}{d_l^\alpha}(c_{n+l}^\dagger c_n^\dagger + c_n c_{n+l})\Big\}.
\end{split}
\end{equation} 
where $t>0$ is the hopping amplitude, $\mu$ is the chemical potential,
$\Delta=|\Delta|e^{i\Theta}$ is known as the complex  superconducting gap and $c_n$'s ($c^\dagger_n$'s) are the spin polarised fermionic annihilation (creation) operators defined at every site $n$ of the chain with total sites $N$. The superconducting pairing term being a function of the distance $d_l=\text{Max}[l,L-l]$ between any two sites in the lattice is long range interacting with the strength of interaction decaying with a decay exponent $\alpha > 0$. Although the total fermionic number is not conserved, the parity operator (total fermionic number modulo two) commutes with the Hamiltonian and is conserved.

Focussing on  the rather well-known  short-range limit ($\alpha \to \infty$) \ct{kitaev01}, when the system is in the topological phase, there are two Majorana modes (MM) at each end of the open chain. The two MMs having the same degrees of freedom as an ordinary fermion can either be together occupied or unoccupied. Since the energy of the MMs are zero, these two possible states (occupied/unoccupied) are both ground states thereby rendering the ground state  of a short-range  1D Kitaev chain two fold degenerate with different parities: (i) a bulk with even fermion parity and unoccupied MMs, (ii) while populating the two Majorana modes at the edges (in addition to the bulk) amounts to a single ordinary fermion and odd parity. \ {As we shall illustrate below, the LRK chain is {\it topologically}
short-ranged when the decay parameter $\alpha >3/2$.}

Throughout the rest of the paper, without the loss of generality, we will set $t=\Delta=1/2$ and assuming periodic boundary conditions, one can implement a Fourier transformation to rewrite the Hamiltonian in the Nambu spinor basis, $\psi_k=(c_k,c^\dagger_{-k})^T$. The thermodynamic limt of $N\to\infty$ yields \ct{vodola14,vodola16,viyuela16},
\begin{equation}
H = \int_0^{2\pi}\frac{\operatorname{d}\!k}{2\pi} \Psi_k^\dagger H_k \Psi_k,
\end{equation}
where
\begin{equation} 
H_k = -  f_\alpha(k)\sigma_y - \left(\mu + \cos{k}\right)\sigma_z.
\end{equation}
and 
\begin{equation} \label{eq_polylog}
f_\alpha(k)=\sum_{l=1}^{N-1}\frac{\sin{(kl)}}{l^\alpha}
\end{equation}

The eigenvalues of this Hamiltonian are 
\begin{equation} \label{eq_spectrum}
E_k^\pm = \pm \sqrt{\left( \mu +\cos{k}\right)^2 + 
\left(  f_\alpha(k)\right)^2}. 
\end{equation}
Moreover, to simplify matters, we consider the Hamiltonian in a rotated basis, with the Bloch vectors of the Hamiltonian lying on the equatorial plane so that  $H_k$
assumes the form,
\begin{equation}
H_k = - \frac{\Delta_k}{2}\vec{n}_k\cdot \vec{\sigma}, 
\end{equation} 
where $\vec{\sigma} = (\sigma_x, \sigma_y, \sigma_z)$ are the Pauli matrices and 
\begin{align}
\vec{n}_k &= \frac{2}{\Delta_k}\left( \mu+\cos{k},f_\alpha(k),0\right),\\ 
\Delta_k &= 2|E_k^\pm|.
\end{align}

It is noteworthy that the LRK chain is classified under the BDI symmetry class of topological insulators and superconductors \cite{schnyder08,kitaev09} and is particle-hole, time-reversal, and chiral symmetric. These symmetries restrict the movement of the Bloch vector $\vec{n}_k$ from the sphere $S_2$ to the circle $S_1$ on the $x-y$ plane resulting in a mapping from the Hamiltonians $H_k$ on the Brillouin zone (BZ) $k\in S_1$ onto the winding vectors $\vec{n}_k\in S_1$. This mapping yields a topological $Z_2$ invariant called the winding number $\omega$, which is the angle (modulo $2\pi$) subtended by $\vec{n}_k$ when quasi-momentum $k$ is varied across the BZ from $-\pi$ to $\pi$, where:
\begin{equation} \label{eq_winding}
\omega=\frac{1}{2\pi}\oint{\frac{\partial_k n^y_k}{n^x_k} \operatorname{d}\!k}
\end{equation}
Alternatively, one can consider an adiabatic transport of the system from a certain crystalline momentum via a reciprocal lattice vector. The eigenstate of the lower band of the system $|g_k\rangle$ then picks up a Berry/Zak phase $\phi_Z$  \ct{berry84,aharonov87,zak89} which is generally quantized ($0$ or $\pi$) and has a one-to-one correspondence with the winding number Eq.~\eqref{eq_winding} defined above.
\begin{equation}
\phi_Z=i\oint{\langle g_k|\partial_k|g_k\rangle \operatorname{d}\!k}
\end{equation}\\

\begin{figure}
\includegraphics[width=9cm,height=8cm]{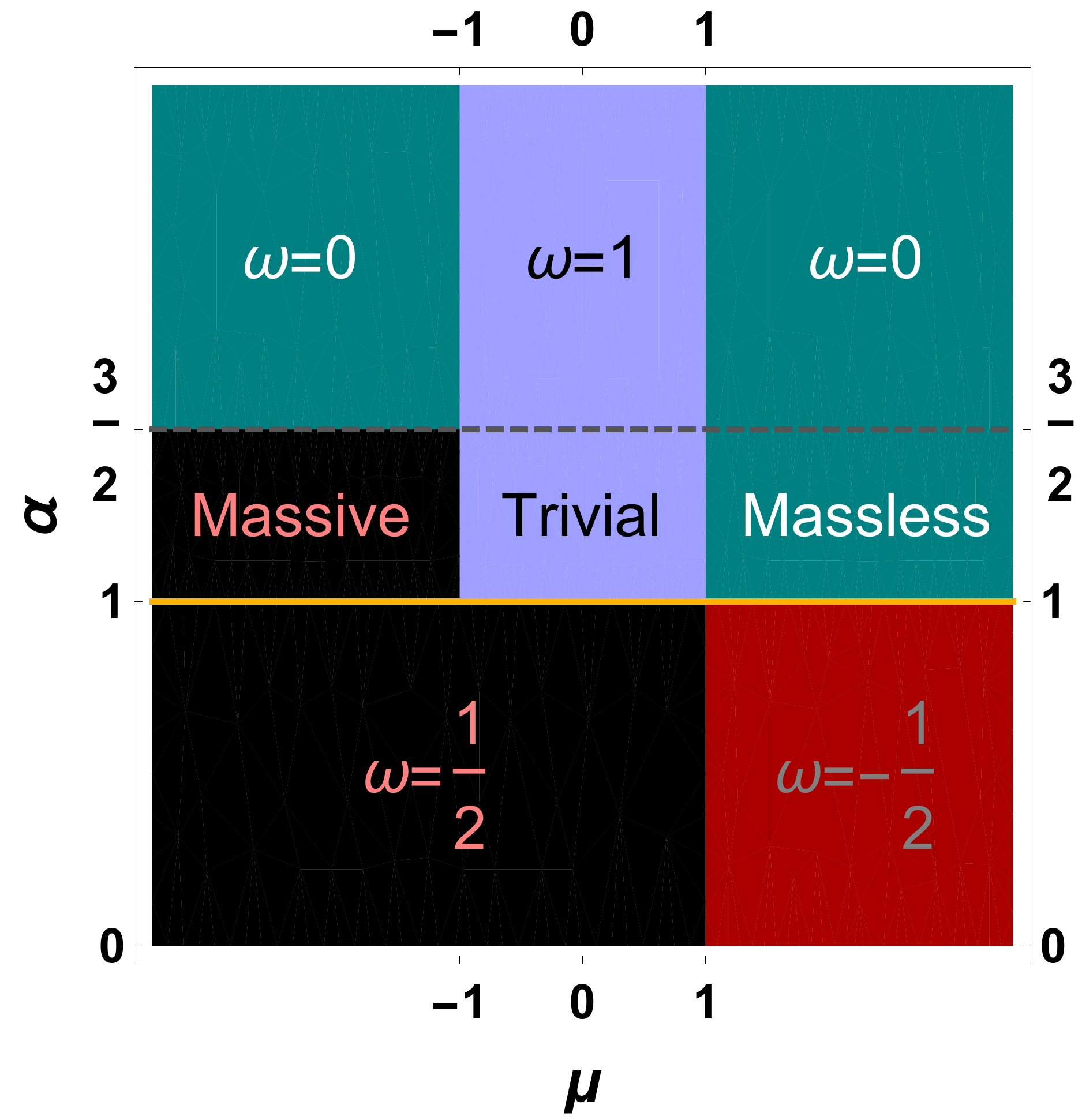}
\caption{Phase diagram of the LRK chain in the $\mu - \alpha$ plane: for $\alpha>3/2$ the phase diagram of this model is topologically equivalent to the short range Kitaev chain, whereas for $\alpha<1$, the model hosts massive Dirac edge modes for $\mu<1$ and is characterised by half-integer winding number. There is a crossover phase in between (for $1<\alpha<3/2$) with no well-defined winding number.}
\label{PD}
\end{figure}

In the thermodynamic limit, the polylogarithmic function $f_\alpha (k)$ in Eq.~\eqref{eq_polylog} that encodes all the information about the long-range pairing, is divergent at $k=0$ for $\alpha < 1$ and this results in the likewise divergence of the dispersion relation (see Eq.~\eqref{eq_spectrum}) and the group velocity $(\partial E_\pm(k)/\partial k)$. Moreover, the impossibility in gauging away the divergence from $k=0$ generates a topological singularity. Therefore,  according to the behavior of $f_\alpha (k)$ at $k=0$, the existence of three different topological sectors depending on the exponent $\alpha $ have been rigorously established by Viyuela $et~al.$ \cite{viyuela16}:\\

a) The $\alpha > 3/2$ sector also known as the Majorana sector is equivalent to the topological phase of the short-range Kitaev chain \cite{kitaev01}. The $|\mu| > 1 $ phase is topologically trivial and is marked by the absence of the Majorana zero modes (MZMs). On the other hand, in the region $\mu \in \left(-1,1\right)$, the MZMs are ever present [see Fig.~\ref{PD}]. The presence of a $U(1)$ phase discontinuity at $k=0$ in the eigenvector $|g_k\rangle$ and the function $f_\alpha (k)$ being not divergent yields the $Z_2$ invariant $\omega=\frac{\phi_Z}{\pi}=1$ which characterises this phase.\\

b) The $\alpha < 1$ sector is truly an emergent feature of the long range nature of the hopping and is absent in the conventional 1D Kitaev model. In this sector, for $\mu > 1$ the system under open boundary condition is in a trivial phase, with no edge states while, for $\mu < 1$ this system hosts topological massive Dirac fermion at the edges, as shown in the wave function plot in Fig.~2(b) of [\onlinecite{viyuela16}]. This massive Dirac mode (MDM) appears solely due to the coupling induced between the two MZMs at the two distant edges due to the presence of long range superconducting pairing and thus, the MDM formed is highly non-local. Moreover, the MDM although massive is still topological and is thereby protected by the bulk gap. Furthermore, as the ground state of the system in this phase still retains its even parity, populating the MDM which is the first excited state of the system would now require a change in the fermionic parity from even to odd. Therefore, this highly non-local topological quasiparticle is also protected by the fermionic parity. Since no discrete symmetry has been broken, due to the inclusion of the long-range pairing, the system still belongs to the BDI symmetry class. The winding number $\omega$ however, is modified by the topological singularity at $k=0$. This happens because at $k=0$ the adiabatic condition breaks down since both the energy dispersion relation $E_k^{\pm}$ in Eq.~\eqref{eq_spectrum} and the quasiparticle group velocity $\partial_k E^{\pm}_k$ diverge as the Berry/Zak phase $\omega=\phi_Z/\pi$ evolves under parallel transport. For the trivial phase $\mu > 1$, the winding number is $\omega = -1/2$, whereas for the massive Dirac fermion hosting topological phase when $\mu < 1$, it turns out to be $\omega = +1/2$. Although the topological invariant is half-integer, the difference of one unit exists between the two topologically different phases indicating that a topological phase transition separates the two half-integer quantised topological phases.\\

c) The third sector for $\alpha \in (1,3/2)$ not only hosts MZMs for $-1<\mu < 1$ but also include MDM for $\mu >1$. The dispersion relation $E_k^{\pm}$ is no longer divergent, however the group velocity $\partial_k E_k^{\pm}$ is still singular at $k=0$, and hence, a winding number cannot be defined for such a crossover sector. 

In the rest of the article we are only going to focus on the sectors a) and b) with well defined winding numbers, to see how the topological invariant behaves when the chain is in constant contact with a thermal bath at temperature $T$, and is 
described by the Gibbs' state 
\begin{equation} 
\rho (k) = \frac{e^{-H_k/T}}{\mbox{Tr}e^{-H_k/T}} = 
\frac{1}{2}\left( \mathbbm{1} + \tanh{\frac{\Delta_k}{2T}}  
\vec{n}_k\cdot \vec{\sigma}\right).
\label{eq_rho}
\end{equation}

An important question that has been asked now is whether it is possible that a geometric phase factor can also be defined for mixed states, analogous to the Berry/Zak phase for pure states, which can serve as a topological invariant describing the phase structure of 1D chains at finite temperature. The work by Viyuela $et~ al$ \ct{viyuela14} suggests that the Uhlmann geometric phase \ct{uhlmann86,uhlmann89} can play this role. Alternatively, the behavior of interferometric geometric phase for mixed states defined introduced in the Ref. [\onlinecite{sjoqvist00}] have also been studied by Andersson $et~ al.$ \ct{andersson16} for the short-range 1D Kitaev chain and was shown to be a candidate approach. In this work, we illustrate that for 1D Kitaev chain for a long range pairing, although the interferometric approach still manages to capture the topological properties of the pure state, the Uhlmann approach fails miserably even  to reproduce the correct  pure state limit.

\section{ The Uhlmann approach}\label{UA}
Uhlmann's approach \ct{uhlmann86,uhlmann89} is based upon considering pure states in the extended Hilbert space $\HH_A\otimes\HH_B$
which forms the total space of a fiber bundle over the mixed states on $\HH_A$. Now a purification is performed such that:
\begin{equation}
\rho=\Tr_B\left[|\psi\rangle\langle \psi |\right]=ww^{\dagger}
\end{equation}
where the trace is over the auxiliary space $\HH_B$. This description contains a $U(N)$ gauge freedom since under $w\to wU(N)$, $\rho\to wU(N)U^{\dagger}(N)w^{\dagger}$ remains unchanged.
A geometric phase can be associated to any curve in the base manifold once a parallelism condition for curves in the total space is defined. A lift $w(k)$ of $\rho(k)$ is said to be parallel if for every infinitesimal $\delta k$ the probability for the transition from $\psi(k)$ to $\psi(k+\delta k)$ is identical to the fidelity of $\rho(k)$ and $\rho(k+\delta k)$, 
\begin{equation}
\left|\Tr(w(k)^\dagger w(k+\delta k))\right|^2=\Tr\sqrt{\rho (k)^{1/2}\rho(k+\delta k)\rho (k)^{1/2}}.
\end{equation}

The parallelism condition is eventually described in terms of a connection $\A$ \ct{uhlmann91},  which along the velocity fields of square root lifts \ct{hubbner93}, i.e., $w(k)=\sqrt{\rho(k)}$ becomes:

\begin{equation} 
\mathcal{A}\left(\partial_k w\right) = \sum_{i,j}|u_i\rangle \frac{\langle u_i| 
[\partial_k w,w]|u_j\rangle }{p_i+p_j}\langle u_j|, 
\end{equation}
where the $p_i$ and the $\ket{u_i}$ are the eigenvalues and eigenstates of $\rho$ in Eq.~\eqref{eq_rho} and  
\begin{equation} 
p_+ = \frac{1}{2}\big(1+\tanh{\frac{\Delta_k}{2T}}\big),\qquad 
p_- = \frac{1}{2}\big(1-\tanh{\frac{\Delta_k}{2T}}\big).
\end{equation}
The above formula simplifies for a two level system into
\begin{eqnarray}
 \mathcal{A}(\partial_k w) &=& (\sqrt{p_+}-\sqrt{p_-})^2\Big\{|u_+\rangle 
\langle u_+|\partial_k u_-\rangle \langle u_-| \nonumber\\
&+& |u_-\rangle \langle u_-|\partial_k u_+\rangle 
\langle u_+|\Big\}.
\end{eqnarray}
where 
\begin{equation}
\ket{u_+} = \frac{1}{\sqrt{2}}\begin{pmatrix} 1 \\ e^{i\phi} \end{pmatrix},\qquad \ket{u_-} = \frac{1}{\sqrt{2}}\begin{pmatrix} 1 \\ -e^{i\phi} \end{pmatrix}. 
\end{equation}
and $\phi = \arctan{\left(n_y/n_x\right)}$.
As the connection in our case becomes Abelian, 
\begin{equation} 
\mathcal{A}(\partial_k w) = \frac{i}{2}(\partial_k\phi)(\sqrt{p_+}-\sqrt{p_-})^2
\begin{pmatrix}
-1 & 0 \\ 0 & 1
\end{pmatrix},
\end{equation}
path ordering is automatically taken care of in computing 
\begin{equation} 
U = \exp\left( -\oint\!\operatorname{d}\!k\, \mathcal{A}(\partial_k w)\right) 
= \begin{pmatrix}
e^{iB} & 0 \\ 0 & e^{-iB}
\end{pmatrix}, 
\end{equation}
where 
\begin{equation} 
B = \frac{1}{2}\oint\!\operatorname{d}\!k\,(\partial_k \phi)(\sqrt{p_-}-\sqrt{p_+})^2. 
\end{equation}
Let us now remark that even though $\phi(k)$ is periodic, the function $B$ need not be periodic.
Finally, we obtain the Uhlmann's geometric phase as the argument of the phase factor of the function 
\begin{equation}
\begin{split}
&\Tr\left(w(0)^\dagger w(0)U\right)\\
&
= \frac{1}{2}\Big(\sqrt{p_+(0)}+\sqrt{p_-(0)}\Big)^2\cos{B}\,+
\\  &+ \frac{1}{2}\Big(\sqrt{p_1(0)}-\sqrt{p_2(0)}\Big)^2
\cos{\left(\phi(0)+B\right)}. 
\end{split}
\label{eq_trace1}
\end{equation}
Using $\phi(0)=0$, the above equation \eqref{eq_trace1} reduces to
\begin{equation}
\begin{split}
\Tr\left(w(0)^\dagger w(0)U\right)&=\cos(B)\\
&=\cos\left[\frac{1}{2}\oint\!\operatorname{d}\!k\,(\partial_k \phi)\left\{1-\sech\left(\frac{\Delta_k}{2T}\right)\right\}\right]
\end{split}
\label{eq_trace2}
\end{equation}
Let us first study the $T\to 0$ limit. In this limit the Berry/Zak phase for the pure state case (see the phase diagram \ref{PD}) should be: 
\[ \phi_Z=\begin{cases} 
      0 & |\mu|>1~\text{and}~\alpha>\frac{3}{2} \\
      \pi & -1<\mu<1~\text{and}~\alpha>\frac{3}{2} \\
      \text{not applicable} & \forall\mu~\text{and}~1<\alpha<\frac{3}{2} \\
      -\frac{\pi}{2} & \mu>1~\text{and}~1<\alpha\leq 0 \\
      \frac{\pi}{2} & \mu<1~\text{and}~1<\alpha\leq 0
   \end{cases}
\]
In the $T\to 0$ limit, Eq. \eqref{eq_trace2} becomes
\begin{equation}
\begin{split}
\Tr\left(w(0)^\dagger w(0)U\right)
&=\lim_{T\to 0}\cos\left[\frac{1}{2}\oint\!\operatorname{d}\!k\,(\partial_k \phi)\left\{1-\sech\left(\frac{\Delta_k}{2T}\right)\right\}\right]\\
&=\cos\left[\frac{1}{2}\oint\!\operatorname{d}\!k\,(\partial_k \phi)\right]
\end{split}
\end{equation}
on the other hand,
\begin{equation}
\phi_U=\text{Arg}\left[\Tr\left(w(0)^\dagger w(0)U\right)\right]
\end{equation}
reduces to 
\begin{equation}
\phi_U=\text{Arg}\left[\cos\left(\frac{1}{2}\oint\!\operatorname{d}\!k\,(\partial_k \phi)\right)\right]
\end{equation}
This yields:
\[ \phi_U=\begin{cases} 
      0 & |\mu|>1~\text{and}~\alpha>\frac{3}{2} \\
      \pi & -1<\mu<1~\text{and}~\alpha>\frac{3}{2} \\
      \text{not applicable} & \forall\mu~\text{and}~1<\alpha<\frac{3}{2} \\
      \text{undefined} & \mu>1~\text{and}~1<\alpha\leq 0 \\
      \text{undefined} & \mu<1~\text{and}~1<\alpha\leq 0
   \end{cases}
\]\\
We observe that although the Uhlmann phase $\phi_U$ equals the Berry/Zak phase $\phi_Z$ in the pure state limit for all range of $\mu$\ when $\alpha>1$, $\Tr\left(w(0)^\dagger w(0)U\right)=0$ for all $\mu$ in the strong long range  limit when $\alpha<1$ results in $\phi_U$ being undefined. Therefore, one of the key results of our work is that for the 1D Kitaev chain with long range hopping, the Uhlmann phase fails to detect the topological phase transition at $\mu=1$ for $\alpha<1$ in the pure state limit. It is therefore necessary to resort to a different geometric approach which with a well defined pure state limit can predict the fate of the topological phases when the system is described by mixed quantum states.

\section{ The interferometric phase}\label{IGP}
In the geometric interferometric phase approach by Sjoqvist et al, a normalised state, under purification is represented by $|w\rangle \in \HH_w$ where $\HH_w=\HH_S\bigotimes\HH_A$, $\HH_S$ is the Hilbert space of the system, $\HH_A$ is the Hilbert space spanned by ancillary states and
\begin{equation}
|w\rangle = \sum_i {\sqrt{p_i}|\psi_i\rangle\bigotimes |\psi^{'}_i\rangle}
\end{equation}
with $|\psi^{'}_i \rangle \in \HH_A$ and the index $i$ runs over the dimensions of the Hilbert space $\HH_S$ (or $\HH_A$). Therefore, the original density matrix is obtained by tracing over the ancillary states:
\begin{equation}
\rho=\Tr_A\left(|w\rangle\langle w|\right)
\end{equation}
Let the states $|w(k)\rangle$ be parametrized by a continuous parameter $k$, with $|w(k)\rangle$ tracing out a curve in the Hilbert space $H_w$. A metric is defined in $H_w$ as the measure of distance between two states as $d = |||w(k_1)\rangle - |w(k_2)\rangle||$. Let us note that the two states $|w(k_1)\rangle$ and $|w(k_2)\rangle$ are said to be parallel if the distance between them is minimum. But, the purification
states $|w(k)\rangle$ also have a phase ambiguity or a U(1) gauge freedom as under a gauge transformation $|w(k)\rangle \to e^{i\delta(k)}|w(k)\rangle$ produces the same density matrix and preserves inner products in the space $\HH_w$, which needs to be fixed to generate a unique trajectory in $\HH_w$. This gauge fixing is implemented by demanding that two infinitesimally separated states in $\HH_w$ are parallel to each other. We should also note that under such a parallel transport the state of the system $|\psi(k)\rangle$ is only affected while the ancillary states $|\psi^{'}(k)\rangle$ are not. For the purifications, using the orthonormality of $|w(k)\rangle$, the corresponding parallel transport condition can be recast  to the form:
\begin{equation}
\langle w(k)|\partial_k w(k)\rangle = \Tr\left(\rho(0)V^{\dagger}(k)\partial_k V(k)\right) = 0
\end{equation}
where $\rho(0)=\sum_i p_i |\psi_i(0)\rangle \langle \psi_i(0)|$ and 
\begin{equation}
V(k)=e^{-\int_0^k \!\operatorname{d}\!k^{'}\,\langle\psi(k^{'})|\partial_{k^{'}} \psi(k^{'})\rangle}
\end{equation}\\
In summary, if we consider a family of density operators  parametrised by $k$,
\begin{equation}
\rho(k)=\sum_ip_i(k)|\psi_i(k)\rangle\langle\psi_i(k)|
\end{equation}
such that for each $k$, the eigenvalues $p_i(k)$ are non-degenerate, the parallel gauge fixing condition,
\begin{equation}
\langle\psi_i(k)|\partial_k \psi_i(k)\rangle=0
\end{equation} 
after a parallel transport across the whole 1D Brillouin zone, yields the interferometric phase \ct{andersson16} of $\rho(k)$:
\begin{equation}
\theta_g=\text{Arg}\left[\sum_i \sqrt{p_i(0)p_i(2\pi)} \langle\psi_i(0)|V_i(2\pi)|\psi_i(2\pi)\rangle\right]
\end{equation}
where $V_i(2\pi)=e^{-\oint \!\operatorname{d}\!k^{'}\,\langle\psi_i(k^{'})|\partial_{k^{'}} \psi_i(k^{'})\rangle}$.\\
It is now straightforward to calculate this interferometric phase $\theta_g$ in the case of the 1D Kitaev model with long range hoppings which is essentially a two-level quantum system for each independent $k$-mode. Using Eqs. (15), (8) and (5) and identifying $\psi_i(k)$ as $u_i(k)$ in Eq. (17), we finally obtain:
\begin{eqnarray}
\theta_g &=&\text{Arg}\left[\exp\left(-\frac{i}{2}\oint \!\operatorname{d}\!k^{'}\,\frac{\partial \phi}{\partial k^{'}}\right)\sum_{i=\pm}p_i(0)\right]\\
&=&\frac{1}{2}\oint \!\operatorname{d}\!k^{'}\,\frac{\partial \phi}{\partial k^{'}}=\phi_Z
\end{eqnarray}
It can now easily be seen that not only the interferometric geometric phase $\theta_g$ reduces to the Berry/Zak phase in the pure state limit for all values of $\alpha$  and thus reproduces the phase diagram properly and also  $\theta_g$ is completely independent of the temperature of the bath. This happens as the phase accumulated by both the eigenstates under parallel transport across the $k$-space remains the same and is identical to the Berry/Zak phase $\phi_Z$.

\section{Discussion and concluding comments:}\label{sec_conclusion}
The 1D Kitaev chain with short range (nearest neighbour) superconducting pairings is a model of p-wave topological superconductor \ct{kitaev01} which posses a topological phase characterised by a $Z_2$ topological invariant in the zero temperature limit. After coupling this model to a bath maintained at a constant temperature $T$, it's topological behavior has been thoroughly investigated in the works of Viyuela $et~al.$ \ct{viyuela14}and Andersson $et~al.$ \ct{andersson16}. While the former works used the Uhlmann phase approach to provide an order parameter, the latter resorted to the geometric interferometric phase to ascertain its  topological aspects as both these approaches correctly reproduce the pure state topological nature of this model. The Uhlmann phase approach predicts the presence of a critical temperature $T_c$ beyond which the system loses it's topological behaviour. But it also has a memory effect which prevents it from determining the fate of the edge modes at finite temperatures. The interferometric phase on the other hand does not detect any phase transition in temperature, but it correctly captures the zero temperature phase portrait of this model. 

In our work, we have considered a generalised version of  1D Kitaev chain \ct{vodola14} with a superconducting pairing which is now long ranged. The phase diagram of this model is different as it hosts a new massive Dirac phase characterised by a half integer winding number and is the sole result of the long ranged nature of the superconducting term. Having prepared the state of this system in a (mixed) Gibbs' state which is in thermal equilibrium with a bath at finite temperature $T$ the effect of the long ranged nature of the interaction on the topological behaviour is probed using both the aforementioned geometric approaches. We interestingly observe, that the Uhlmann phase approach in the extreme long range limit ($\alpha <1$) fails to detect the zero temperature behaviour of this model. On the other hand, the interferometric phase approach although correctly reproduces the pure state topological limit, invariably fails to capture any topological phase transition with temperature. \  {Our study, therefore, establishes that both the Uhlmann and interferometric phase approaches are inadequate in
describing the finite temperature topology of a LRK chain.}

\section*{ACKNOWLEDGEMENTS}
A.D. acknowledges financial support from SERB, DST, India.

\end{document}